\documentclass[onecolumn,showpacs,preprintnumbers,amsmath,amssymb]{revtex4-2}
\usepackage{graphicx}
\usepackage{dcolumn}
\usepackage{bm}
\usepackage{rotating}
\usepackage{lipsum}
\usepackage{xcolor}
\usepackage{tikz}

    {\par\small\addvspace{4.5ex plus 1ex}%
     \vskip -\parskip
     \ifx\relax#1\relax
        \def\@decl@date{}%
     \else
        \def\@decl@date{\NEWfeature{#1}}%
     \fi
     \noindent\hspace{-\leftmargini}%
     \begin{tabular}{|l|}\hline\ignorespaces}%
    {\\\hline\end{tabular}\nobreak\@decl@date\par\nobreak
     \vspace{2.3ex}\vskip -\parskip}

\def\Red{} 
\def\Black{} 

\def\lpybox#1{} 
\def\lpybox#1{\fbox{\Red{\bf#1}\Black}} 

\begin{document}
\title{Charge-state dynamics of barium ions in high-pressure xenon and its implications for Barium-Tagging in $0\nu\beta\beta$ searches}
\author{A. Peralta Conde}
\affiliation{Universidad Internacional de la Rioja (UNIR). www.unir.net. Spain.}
\email{alvaro.peralta@unir.net}
\begin{abstract}
Barium tagging (BaTa) is one of the most promising techniques for achieving a nearly background-free search for neutrinoless double-beta decay ($0\nu\beta\beta$) in high-pressure xenon time detection chambers. However, the experimental implementation of BaTa depends critically on the chage-state dynamics of the daughter Ba$^{2+}$ ion produced in the nuclear decay event. In this work, I review the possible recombination channels and evaluate their physical viability. The obtained results indicate that although binary recombination channels are strongly suppressed, three-body recombination assisted by neutral xenon atoms constitutes a physically plausible mechanism for the conversion of Ba$^{2+}$ into Ba$^+$ on timescales -milliseconds- comparable to the characteristic detection times in the NEXT experiment. These results suggest that the barium charge state should be regarded as a dynamical quantity with direct implications for the design and experimental implementation of BaTa techniques.
\end{abstract}
%
\maketitle
\section{Introduction}

Neutrinos remain among the least understood particles in the Standard Model. Despite the remarkable experimental progress achieved over the last decades, some of their most fundamental properties remain unknown. In particular, determining the neutrino mass ordering and establishing whether neutrinos are Dirac or Majorana particles are among the main open questions in particle physics. Their answers would have profound implications not only for our understanding of the Standard Model but also for cosmology, especially regarding the origin of the matter-antimatter asymmetry through leptogenesis (see, for example, \cite{Athar22} and references therein).

One of the most promising experimental approaches to determine whether neutrinos are Majorana or Dirac particles relies on the search for neutrinoless double beta decay ($\beta\beta0\nu$). Double beta decay is a second-order weak nuclear process in which a nucleus with atomic number $Z$ transforms into a nucleus with atomic number $Z+2$ while preserving its mass number $A$. Two different channels are possible for this transition. In the standard two-neutrino mode ($\beta\beta2\nu$ mode) two neutrons are converted into two protons with the emission of two electrons and two electron antineutrinos ($(Z, A)\rightarrow(Z+2, A)+2e^-+2{\bar{\nu}}_e$). This process is allowed in the Standard Model and, although it has a long lifetime, resulting in a low probability of occurrence, it has been experimentally observed (see for example the results of the KamLAND-Zen collaboration, \cite{Gando13} and references therein). In contrast, the neutrinoless mode ($\beta\beta0\nu$ mode), i.e., the process $(Z, A)\rightarrow(Z+2, A)+2e^-$, has never been observed despite the considerable experimental efforts. Its predicted lifetime, longer than $10^{26}$\,years \cite{Gando16}, makes its observation an experimental challenge and one of the major goals for contemporary particle physics. Its observation would imply the violation of lepton-number conservation and would constitute direct evidence that neutrinos are Majorana particles \cite{Majorana37}.

Among the experiments devoted to this search, the Neutrino Experiment with a Xenon TPC (NEXT), located at the Canfranc Underground Laboratory (LSC), employs high-pressure gaseous $^{136}$Xe to search for neutrinoless decay by reconstructing the energy and topology of the emitted electrons (see, for example, \cite{Novella23} and references therein). Since the expected half-life of the process exceeds $10^{26}$ years, if it exists, the expected signal rate is extremely low. Hence, achieving the required experimental sensitivity demands a combination of large detector masses, long exposure times, excellent energy resolution, and operation in an ultra-low-background environment based on underground laboratories and highly radiopure detector components. Even under these conditions, the expected signal from the $\beta\beta0\nu$ mode is extremely low, making further background suppression essential for extending the experimental sensitivity, and ensuring that no false-positive background events are identified as signal events. 

A key strategy to reduce the experimental background is Barium Tagging (BaTa). The idea consists of correlating the detection of the emitted electrons with the identification of the daughter barium atom produced in the nuclear transition $^{136}\mathrm{Xe} \rightarrow^{136}\mathrm{Ba}^{2+} + 2e^- + (2\bar{\nu}_e)$. Since barium can only be produced by the decay of $^{136}$Xe, its unambiguous identification would provide an essentially background-free signature of the event. Because of its enormous potential for background rejection, BaTa has attracted considerable attention during the last decades, both for liquid- and gaseous-xenon detectors \cite{Moe91,Danilov00,Sinclair11,Mong15,Albert15,Nygren15,Thapa19,nEXO19}.

However, the practical implementation of BaTa critically depends on the charge state of the daughter atom after the nuclear decay. Different experimental approaches rely on different assumptions regarding the charge evolution of the produced barium. If the ion survives as Ba$^{2+}$, Rivilla \textit{et al.} \cite{Rivilla20} demonstrated the feasibility of using a Fluorescent Bicolor Indicator (FBI) capable of selectively capturing doubly ionized barium. On the other hand, if the ion rapidly captures one or two electrons, becoming Ba$^+$ or neutral Ba, laser-induced fluorescence techniques constitute an attractive alternative. In particular, Peralta \cite{Peralta25} proposed exploiting the characteristic atomic transitions of barium to identify its presence without requiring molecular tagging agents and, therefore, considerably simplifying the experimental setup.

Despite the importance of the charge state of the daughter atom for the design of future BaTa systems, comparatively little attention has been devoted to estimating whether the doubly charged ion can actually survive in the high-pressure xenon environment immediately after the nuclear decay. Most of the proposed experimental techniques implicitly assume one charge state or another, but quantitative estimates of the survival of the Ba$^{2+}$ ion remain scarce. Consequently, it is still unclear under which physical conditions each of the currently proposed tagging strategies is expected to be feasible.

In this work I address this question by developing a simplified physical model to estimate the survival probability of the Ba$^{2+}$ ion in such a high-pressure environment. Rather than attempting a complete microscopic description of the thermalization, charge-exchange and recombination processes, which would require a considerably more sophisticated treatment and lies beyond the scope of the present work, I derive physically motivated bounds for the survival of the doubly charged ion. Although approximate, these estimates provide useful guidance for assessing the experimental viability of the different BaTa strategies.

\section{Dynamical estimate of the Ba$^{2+}$ survival probability in high-pressure xenon}

In this section, a simplified physical picture is constructed for the microscopic environment created immediately after the double beta decay of $^{136}$Xe in high-pressure xenon gas. Rather than attempting to predict the lifetime of the daughter ion quantitatively, the purpose is to identify the characteristic spatial, temporal and density scales that determine whether Ba$^{2+}$ survives long enough to be detected or recombines to Ba$^{+}$, or eventually to neutral Ba, on the timescales relevant to the NEXT experiment.

In a double beta decay event, the daughter nucleus is produced together with a doubly charged barium ion and two energetic electrons whose total kinetic energy can reach $Q_{\beta\beta}\simeq2.46$ MeV. These electrons propagate through the xenon gas with no preferred direction, losing energy predominantly through inelastic collisions with neutral xenon atoms. Under the operating conditions for the NEXT-100 experiment, 15\,bar of xenon at room temperature, collisional losses largely dominate over radiative processes, and the stopping power of MeV electrons is well described by the Bethe formalism \cite{Segre53}.

For the recombination problem, however, the entire ionization track is not equally relevant. The energetic $\beta$ electrons emitted in the decay simply create the ionization track, whereas the electrons released during these ionization processes, commonly known as $\delta$-electrons, constitute the only population that can eventually recombine with the daughter Ba$^{2+}$ ion. Since the charges produced far from the decay vertex are rapidly separated by the external drift field, only the first ionization events occurring sufficiently close to the decay point are expected to contribute to the local electron environment surrounding the ion. Consequently, the characteristic distance to the first ionizing collision provides a natural estimate of the spatial region where recombination may occur.

The electron-impact ionization cross section in xenon depends sensitively on the kinetic energy of the incident electron. It reaches a maximum value of the order of $\sigma_{\mathrm{ion}}\sim5\times10^{-18}\,\mathrm{cm^{2}}$ for electron energies of a few tens of electronvolts and decreases at higher energies \cite{Sorokin00,NIST14}. For the $\sim1$ MeV $\beta$ particles produced in double beta decay, a representative value is $\sigma_{\mathrm{ion}}\sim10^{-18}\,\mathrm{cm^{2}}$. Using the xenon atomic density under the operating conditions of the NEXT-100 experiment (15\,bar, $T\simeq300$ K),
\begin{equation}
n_{\mathrm{Xe}}\simeq4\times10^{20}\,\mathrm{cm^{-3}},
\end{equation}
the corresponding mean free path for ionizing collisions can be estimated as
\begin{equation}
\lambda_{\mathrm{ion}}=\frac{1}{n_{\mathrm{Xe}}\sigma_{\mathrm{ion}}}\simeq
25\,\mu\mathrm{m}.
\end{equation}

It is important to emphasize that $\lambda_{\mathrm{ion}}$ does not represent a fixed distance at which ionization occurs, but rather the mean of an exponential probability distribution. Since ionization is a stochastic process, the probability that the first ionizing collision occurs before a distance $x$ is
\begin{equation}
P(x)=1-\exp\!\left(-\frac{x}{\lambda_{\mathrm{ion}}}\right).
\end{equation}
Thus, the probability that the first ionization takes place within one mean free path is
\begin{equation}
P(\lambda_{\mathrm{ion}})=1-e^{-1}\simeq0.63.
\end{equation}
Throughout the following discussion, $\lambda_{\mathrm{ion}}$ is therefore used as the characteristic distance to the first ionization event. Although this approximation neglects the stochastic nature of individual events, it provides an adequate description for the order-of-magnitude estimates developed in this work.

Because the differential ionization cross section strongly favors small energy transfers \cite{Sorokin00,NIST14}, most $\delta$-electrons carry only a few tens of electronvolts of kinetic energy. A representative energy range is
\begin{equation}
E_\delta\sim5-60\,\mathrm{eV},
\end{equation}
while only a small fraction reach energies above 100 eV. The production of keV $\delta$-electrons by MeV primary electrons is therefore strongly suppressed.

The $\delta$-electrons may in turn produce further ionization before thermalizing, giving rise to small ionization sub-cascades. The average number of additional electron-ion pairs generated by a $\delta$-electron of initial energy $E_\delta$ can be estimated as
\begin{equation}
N_{\mathrm{sec}}\simeq\frac{E_\delta}{W_{\mathrm{Xe}}},
\end{equation}
where $W_{\mathrm{Xe}}$ is the average energy required to produce one electron-ion pair in gaseous xenon. This experimentally determined quantity accounts not only for ionization losses but also for excitation and sub-excitation processes. For xenon gas, $W_{\mathrm{Xe}}\simeq21.9$ eV \cite{Bolotnikov97}. Consequently, for the typical $\delta$-electron energies estimated above ($E_\delta\sim5$--60 eV), one expects only
\begin{equation}
N_{\mathrm{sec}}\sim1-3
\end{equation}
additional ionizations. Therefore, each primary ionization event produces a small cluster containing only a few electrons rather than an isolated electron-ion pair.

The characteristic size of these elementary ionization clusters can be estimated from the transport properties of low-energy electrons in xenon. For electron energies between approximately 10 and 100\,eV, the electron-impact ionization cross section is of the order of 
\begin{equation}
\sigma_{\mathrm{ion}}\sim5\times10^{-16}\,\mathrm{cm}^2,
\end{equation}
corresponding to a mean free path of approximately
\begin{equation}
\lambda_{\delta e^-}\sim50\,\mathrm{nm}
\end{equation}
at 15 bar \cite{Sorokin00}. Since a low-energy $\delta$-electron undergoes only a few collisions before thermalizing, the resulting sub-cascade is expected to extend over a characteristic radius of
\begin{equation}
r_{\mathrm{sub}}\sim10-100\,\mathrm{nm},
\end{equation}
which defines the typical size of an elementary ionization cluster. Taking a representative cluster radius of
\begin{equation}
r_c\sim50\,\mathrm{nm},
\end{equation}
the corresponding volume is
\begin{equation}
V_c=\frac{4}{3}\pi r_c^3\simeq5\times10^{-16}\,\mathrm{cm^3}.
\end{equation}
Assuming that such a cluster contains only $\mathcal{O}(1)$ electrons, the corresponding instantaneous electron density is
\begin{equation}
n_e^{(\mathrm{cluster})}\sim\frac{1}{V_c}\sim2\times10^{15}\,\mathrm{cm}^{-3}.
\end{equation}

This instantaneous electron density characterizes the interior of an elementary ionization cluster immediately after its formation. However, it should not be interpreted as the electron density surrounding the daughter Ba$^{2+}$ ion. Before reaching distances comparable to the mean ionization length, electrons undergo numerous collisions with xenon atoms, rapidly losing memory of their initial energies while diffusing away from the ionization cluster. Experimental measurements of the total electron-xenon scattering cross section indicate values in the range \cite{Dababneh82,Hayashi03}
\begin{equation}
\sigma_{e-\mathrm{Xe}}\sim10^{-16}-10^{-15}\,\mathrm{cm^2}
\end{equation}
for electron energies between 1 and 100 eV. Under the operating conditions of the NEXT-100 detector, this corresponds to electron mean free paths of only
\begin{equation}
\lambda_e=\frac{1}{n_{\mathrm{Xe}}\sigma_{e-\mathrm{Xe}}}\sim1-10\,\mathrm{nm}.
\end{equation}
Low-energy electrons therefore undergo many collisions over extremely short distances, efficiently dissipating their kinetic energy through excitation and ionization. The characteristic thermalization time can consequently be estimated as
\begin{equation}
\tau_{\mathrm{th}}\sim10^{-13}-10^{-12}\,\mathrm{s},
\end{equation}
which is several orders of magnitude shorter than the time required to travel distances of tens of micrometers. Consequently, electrons reaching distances comparable to the characteristic separation between the Ba$^{2+}$ ion and the first ionization event are expected to be essentially thermalized. The initially dense ionization cluster therefore evolves into a much more dilute cloud of thermal electrons.

Assuming that the first ionization event occurs at a characteristic distance
\begin{equation}
r_0\simeq\lambda_{\mathrm{ion}}\simeq25\,\mu\mathrm{m},
\end{equation}
a simple geometrical estimate of the local electron density surrounding the daughter ion is
\begin{equation}
n_e\sim\frac{1}{\frac{4}{3}\pi r_0^3}\sim10^7\,\mathrm{cm^{-3}}.
\end{equation}
This estimate intentionally neglects several effects, including Coulomb attraction between the ion and nearby electrons, the external electric field, and electron diffusion during transport. Nevertheless, it provides a reasonable first-order estimate of the microscopic environment in which recombination must occur. The important conclusion is that Ba$^{2+}$ is no longer immersed in the extremely dense electron population initially produced within the ionization cluster, but rather in a much more dilute gas of thermalized electrons. The density and temperature of this electron population are therefore expected to determine the subsequent recombination dynamics.

\subsection{Possible recombination mechanisms}

Once the microscopic environment surrounding the daughter ion has been established, the next step is to identify the physical processes that may convert Ba$^{2+}$ into Ba$^{+}$ or, eventually, into neutral barium. The possible recombination mechanisms can be broadly classified into binary mechanisms, involving only the ion and a single collision partner, and ternary mechanisms, in which a third body is required to dissipate the excess energy and stabilize the recombined ion.

\subsubsection{Binary recombination mechanisms}

The most relevant binary channels are radiative recombination, transient molecular complex formation, and collisional charge transfer.

\begin{itemize}
\item \textbf{Radiative recombination}
\begin{equation}
\mathrm{Ba}^{2+}+e^-\rightarrow\mathrm{Ba}^{+}+h\nu .
\end{equation}
In this process the excess binding energy is released through the emission of a photon. Radiative recombination is generally inefficient in weakly ionized gases because electron capture and photon emission must occur simultaneously. Furthermore, the low electron density estimated above further suppresses the corresponding recombination rate \cite{Smirnov01}. Therefore, radiative recombination is not expected to play a significant role under NEXT operating conditions.

\item \textbf{Formation of transient BaXe molecular complexes}
\begin{equation}
\mathrm{Ba}^{2+}+\mathrm{Xe}\rightarrow(\mathrm{BaXe})^{2+*}\rightarrow\mathrm{Ba}^{+}+\mathrm{Xe}^{+}.
\end{equation}

Collisions between Ba$^{2+}$ and xenon atoms may transiently form molecular complexes capable of redistributing charge. The efficiency of this mechanism depends on the detailed molecular potential-energy surfaces and on the lifetime of the intermediate complex, making its quantitative contribution difficult to estimate without dedicated molecular calculations. However, the efficiency of this pathway depends on specific collision geometries and short-lived molecular configurations, which makes the overall probability of the process relatively small under thermal gas conditions \cite{Smirnov01}.

\item \textbf{Collisional charge transfer}
\begin{equation}
\mathrm{Ba}^{2+}+\mathrm{Xe}
\rightarrow
\mathrm{Ba}^{+}+\mathrm{Xe}^{+}.
\end{equation}
The efficiency of this reaction is governed by the ionization potentials of the two species. The first ionization potential of xenon is $I_{\mathrm{Xe}}=12.13\,\mathrm{eV}$, whereas the second ionization potential of barium is $I_{\mathrm{Ba}^{+}}=10.00\,\mathrm{eV}$. The reaction is therefore endothermic by approximately 2 eV. Since the thermal energy of xenon atoms at room temperature is only $k_BT\simeq0.026\,\mathrm{eV}$, this process is expected to be strongly suppressed under the operating conditions of the NEXT experiment \cite{Janev85}.
\end{itemize}

\subsubsection{Three-body recombination as a possible dominant mechanism}

The discussion above suggests that the binary recombination mechanisms are either energetically unfavorable or intrinsically inefficient under the operating conditions of the NEXT experiment. Under these conditions, three-body recombination assisted by a neutral xenon atom is expected to constitute the dominant pathway for ion recombination
\begin{equation}
\mathrm{Ba}^{2+} + e^- + \mathrm{Xe}
\rightarrow
\mathrm{Ba}^{+} + \mathrm{Xe}.
\end{equation}

In this process, the free electron is first captured into a highly excited Rydberg state of the Ba$^{2+}$ ion. The excess binding energy is subsequently transferred to a third body, a xenon atom, during a collision that stabilizes the recombined ion and prevents immediate autoionization. As a result, the recombined Ba$^{+}$ ion can survive after the collision \cite{Smirnov01, Glosik08}.

The three-body recombination rate per ion can be written as
\begin{equation}
R_3=\alpha_3\,n_e\,n_{\mathrm{Xe}},
\end{equation}
where $\alpha_3$ is the three-body recombination coefficient in units of cm$^{6}$\,s$^{-1}$, $n_e$ is the local free-electron density, and $n_{\mathrm{Xe}}$ is the density of neutral xenon atoms.

Reliable values of $\alpha_3$ are difficult to obtain because they depend on several parameters, including the electron temperature, the electron energy distribution, the ion-neutral interaction potential, and the properties of the third body. To the best of our knowledge, neither experimental measurements nor theoretical calculations are available for the specific Ba$^{2+}$-$e^-$-Xe system under the thermodynamic conditions relevant to the NEXT experiment. Thus, I adopt an order-of-magnitude estimate based on measurements performed in weakly ionized plasmas. Glosik \emph{et al.} \cite{Glosik08} report ternary recombination coefficients for H$_3^+$ ions in He/H$_2$/Ar plasmas of approximately
\begin{equation}
\alpha_3\sim3\times10^{-25}\,\mathrm{cm^{6}\,s^{-1}}.
\end{equation}

Although no quantitative scaling law exists to extrapolate this value to xenon, the considerably larger polarizability of Xe suggests that the corresponding three-body recombination coefficient should not be smaller. We therefore adopt the conservative order-of-magnitude estimate
\begin{equation}
\label{three-body-coeff}
\alpha_3(\mathrm{Ba}^{2+}+e^-+\mathrm{Xe})\sim10^{-25}-10^{-24}\,\mathrm{cm^{6}\,s^{-1}}.
\end{equation}

However, before estimating the recombination time, it is necessary to determine whether a sufficient population of electrons can remain close enough to the Ba$^{2+}$ ion for ternary recombination to occur. The following section therefore examines the local electron dynamics around the daughter ion.

\subsection{Local electron dynamics around the Ba$^{2+}$ ion}

The previous section identified three-body recombination as the most plausible mechanism for the conversion of Ba$^{2+}$ into Ba$^{+}$ or, eventually, into neutral Ba. However, the corresponding recombination rate depends critically on the availability of electrons in the immediate vicinity of the daughter ion. In this section I determine whether thermalized electrons produced in the ionization track can remain close enough to the Ba$^{2+}$ ion for three-body recombination.

As discussed previously, the first ionization events generated by the $\beta$ track occur at distances of the order of $r_0\sim25\,\mu\mathrm{m}$ from the daughter ion. The released $\delta$-electrons produced in this process rapidly lose their excess energy through collisions with xenon atoms. The thermalization time is estimated to be of the order of $\tau_{\mathrm{th}}\sim10^{-13}-10^{-12}\,\mathrm{s}$, after which the electrons can be considered approximately thermalized with the surrounding xenon gas. The estimated distance to the first secondary ionization event significantly exceeds the characteristic length scales of the Coulomb interaction of the  Ba$^{2+}$ ion, raising the question of whether thermal electrons produced in the surrounding ionization cloud can subsequently interact with the daughter ion before being removed by the external drift field.

A useful quantity to characterize this interaction is the Onsager radius, defined as the distance at which the Coulomb potential energy equals the thermal energy of an electron \cite{Onsager38}:
\begin{equation}
r_{\mathrm{O}}=\frac{Ze^2}{4\pi\epsilon_0 k_B T},
\end{equation}
where $Z=2$ for Ba$^{2+}$. At room temperature, this gives $r_{\mathrm{O}}\simeq1.1\,\mu\mathrm{m}$. Within this radius, the Coulomb attraction between the electron and the Ba$^{2+}$ ion exceeds the thermal energy of the electron. Consequently, once an electron enters this region, thermal escape becomes increasingly unlikely, and the probability of repeated close encounters with the ion is enhanced.

A complementary estimate can be obtained by comparing the electric field generated by the Ba$^{2+}$ ion with the electric field employed in the NEXT detector to drift ionization electrons toward the anode, around $E_{\mathrm{drift}}\simeq500\,\mathrm{V\,cm^{-1}}$ \cite{Novella23}. The distance at which the Coulomb electric field produce by the Ba$^{2+}$ becomes comparable with the external drift field can be expressed as:
\begin{equation}
r_*=\sqrt{\frac{1}{4\pi\epsilon_0}\frac{2e}{E_{\mathrm{drift}}}}\simeq2.4\,\mu\mathrm{m}.
\end{equation}

Although the Onsager radius and the field-balance radius originate from different physical considerations, both are of the order of a few micrometers. Although these characteristic lengths are smaller than the estimated distance to the first ionization events, $r_0\sim25\,\mu\mathrm{m}$, they differ by only about one order of magnitude. Consequently, thermalized electrons generated in the surrounding ionization cloud can readily access the Coulomb-dominated region through diffusion. Within this region, neither thermal motion nor the external drift field completely governs the electron dynamics; instead, the attractive Coulomb potential of the Ba$^{2+}$ ion becomes the dominant interaction.

The characteristic diffusion time for a thermal electron to explore a distance $r_0$ can be estimated from the solution of the three-dimensional diffusion equation \cite{Crank79} as:
\begin{equation}
t_{\mathrm{diff}}\sim\frac{r_0^2}{6D_e},
\end{equation}
where $D_e$ is the diffusion coefficient of thermal electrons in high-pressure xenon. Using the representative value $D_e\sim150\,\mathrm{cm^2\,s^{-1}}$, reported for high-pressure xenon \cite{Njoya20}, one obtains
\begin{equation}
t_{\mathrm{diff}}
\sim10^{-9}\,\mathrm{s}.
\end{equation}

This diffusion time is many orders of magnitude shorter than the characteristic times associated with charge transport towards the detector in the NEXT experiment. Therefore, thermalized electrons can rapidly explore the microscopic ionization cloud before significant macroscopic charge extraction takes place. It is important to emphasize that diffusion does not generate a net electron flux towards the Ba$^{2+}$ ion. Instead, it continuously redistributes the thermalized electron population through random motion, allowing electrons to repeatedly sample the surrounding volume. As a result, a fraction of these electrons can enter the Coulomb-dominated region surrounding the daughter ion.

Once an electron reaches distances of the order of the Onsager radius ($r\lesssim r_{\mathrm{O}}$), the attractive Coulomb potential of the Ba$^{2+}$ ion becomes comparable to or larger than both the thermal energy and the force exerted by the external drift field. Consequently, the electron is expected to spend a longer time in the vicinity of the ion than it would in the absence of the Coulomb interaction. This local enhancement of the electron residence time increases the likelihood of electron-ion encounters and provides the physical conditions required for recombination processes to occur. 

\subsection{Estimated recombination time under NEXT conditions}

The previous considerations indicate that a fraction of the electrons generated in the ionization track can remain available for interaction with the daughter Ba$^{2+}$ ion. Under these conditions, three-body recombination assisted by neutral xenon atoms becomes a physically plausible mechanism for modifying the charge state of the daughter ion. The characteristic timescale for this process follows directly from the kinetic equation for ternary recombination in weakly ionized gases \cite{Smirnov01, Raizer91},
\begin{equation}
\tau_3=\frac{1}{\alpha_3n_en_{\mathrm{Xe}}},
\end{equation}
where $\alpha_3$ is the three-body recombination coefficient, $n_e$ is the local density of thermalized electrons available for recombination, and $n_{\mathrm{Xe}}$ is the density of neutral xenon atoms.

Using the electron density estimated in the previous section, $n_e\sim10^{7}\,\mathrm{cm^{-3}}$, together with the xenon density under the operating conditions of NEXT, $
n_{\mathrm{Xe}}\simeq4\times10^{20}\,\mathrm{cm^{-3}}$, the characteristic recombination time is estimated to lie in the range
\begin{equation}
\tau_3\sim0.25-2.5\,\mathrm{ms},
\end{equation}
where the uncertainty reflects the present lack of theoretical or experimental information on the three-body recombination coefficient for the Ba$^{2+}$-$e^-$-Xe system (see Eq.\,\ref{three-body-coeff}) .

It is important to emphasize that this estimate is intentionally conservative. The adopted value of $\alpha_3$ is inferred from measurements in other weakly ionized gases, while the electron density corresponds only to the thermalized electron population available after the initial ionization cascade. Consequently, the calculated recombination time should be regarded as an order-of-magnitude estimate rather than a quantitative prediction.

Nevertheless, the resulting millisecond timescale is comparable to the characteristic times associated with charge transport and event reconstruction in the NEXT experiment. Event reconstruction typically requires several milliseconds following the double beta decay \cite{Francesc}. Therefore, the assumption that the daughter ion necessarily remains in the Ba$^{2+}$ charge state is not supported by this analysis. Instead, the calculations presented here indicate that the conversion
\begin{equation}
\mathrm{Ba}^{2+}\rightarrow\mathrm{Ba}^{+}
\end{equation}
through three-body recombination is a physically plausible process under NEXT operating conditions. 

Furthermore, the present analysis may have important implications for molecular BaTa strategies. Many of the fluorescent molecules proposed for the selective detection of Ba$^{2+}$ possess ionization potentials lower than that of xenon. If such molecules are present in the immediate vicinity of the double-beta decay event, they may be ionized by low-energy secondary electrons that are no longer energetic enough to ionize xenon atoms. As a consequence, these molecular additives could provide an additional source of thermalized electrons in the local environment surrounding the daughter ion, potentially increasing the electron density available for three-body recombination. Although the magnitude of this effect will depend on the molecular concentration, the ionization cross sections, and the electron energy distribution, it suggests that the sensing molecule itself cannot always be regarded as a passive probe. Instead, its presence may modify the microscopic environment in which the charge state of the daughter ion evolves.

\section{Conclusions}

In this work, I present an order-of-magnitude analysis of the possible recombination of the Ba$^{2+}$ ion produced in a double-beta decay event under the operating conditions of the NEXT experiment. The results indicate that three-body recombination constitutes a physically plausible mechanism for the conversion of Ba$^{2+}$ into Ba$^{+}$ on timescales comparable to those relevant for the NEXT detector operation. Although the calculations presented here are intentionally conservative and rely on approximate recombination coefficients, they suggest that the charge state of the daughter barium ion should not be regarded as a fixed property but rather as a quantity that may depend sensitively on the microscopic conditions following the decay. Consequently, recombination effects need to be explicitly considered in the design and implementation of BaTa techniques. The analysis presented in this work provides a physical framework for future theoretical and experimental studies aimed at optimizing BaTa strategies in high-pressure xenon detectors.

\end{document}